\documentclass[aps,twocolumn,prl,showpacs]{revtex4}
\usepackage{bm}
\usepackage{amsbsy}
\usepackage{graphicx}
\newcommand{\beq}{\begin{equation}}
\newcommand{\eeq}{\end{equation}}
\newcommand{\e}{\epsilon}

\begin{document}

\title{Triplet Josephson effect with magnetic feedback}
\author{V.~Braude and Ya.~M.~Blanter}

\affiliation{Kavli Institute of Nanoscience, Delft University of Technology,
2628 CJ Delft, The Netherlands}

\date{\today}
\begin{abstract}
We study AC Josephson effect in a superconductor-ferromagnet
heterostructure with a variable magnetic configuration. The system
supports triplet proximity correlations whose dynamics is coupled to
the magnetic dynamics. This feedback dramatically modifies
the behavior of the junction. The current-phase relation becomes
double-periodic at both very low and high Josephson frequencies
$\omega_J$. At intermediate frequencies, the periodicity in $\omega_J
t$ may be lost.
\end{abstract}
\maketitle

Spin-dependent transport through hybrid structures combining
ferromagnets (F) and normal metals has attracted a lot of interest
in the recent years. This interest is motivated by the
prospect of potential technological applications in the field of
spintronics \cite{spin}. Particular attention is given to
two related effects involving
 mutual influence between the electric current through a structure and
its magnetic configuration.
The first is giant magnetoresistance \cite{Himpsel} in which the conductance
is much larger when different magnetic regions have their magnetic moments
aligned than when they are anti-aligned. The opposite effect is the appearance
of torques acting on magnetic moments when an electric current flows through
the system \cite{Slonczewski}. These non-equilibrium current-induced
torques appear due to non-conservation of spin currents accompanying a
flow of charge through ferromagnetic regions. They
allow manipulation of the magnetic configuration, including
switching between the opposite directions or steady-state precession,
without application of magnetic fields \cite{Katine}.
The two effects combined  promise important
practical applications in nonvolatile memory, programmable logic, and
microwave oscillators.

When the multilayer is coupled to a superconductor (S), an additional
constraint is added, viz. that
the spin current through the superconducting part vanishes
\cite{Brouwer1}. This modifies the non-equilibrium
torques, opening  the possibility of perpendicular alignment of magnetic
moments. A very different
situation arises when a magnetic structure is contacted by {\em two}
superconductors.
In this case, the proximity effect may be present, leading to a
finite Josephson current  through the structure at equilibrium. The
torques generated by this
current correspond to an {\it equilibrium} effective exchange
interaction between the magnetic
moments which can be
controlled by the phase difference between the superconductors
\cite{Brouwer2}. The same mechanism
enables the reciprocal effect in which the supercurrent depends on
the magnetic configuration.

Naive considerations might suggest that the proximity effect
should be suppressed at short distances in the presence of
ferromagnets. However, recently it was shown that a long-range effect
can exist due to triplet  superconducting correlations \cite{prox}.
This triplet proximity effect (TPE), and in particular, the associated
Josephson current, depend essentially on the magnetic configuration of
the system \cite{B-N}. Hence S/F multilayers exhibiting TPE are
especially suitable for
studying the  Josephson-induced magnetic  exchange interaction.

By varying the relative magnetization directions of different magnetic
regions, one can control  the supercurrent flowing through the structure.
Then, if the magnetic configuration is allowed to respond to the
Josephson-current induced torques, it creates feedback for the
supercurrent and considerably modifies it. We show that its  main signature
is frequency doubling in the current-phase relation.

Below, we consider this feedback for a Josephson junction biased by a
dc voltage.
In the AC Josephson effect, the time dependence is normally
determined by the current-phase relation. TPE in  diffusive systems
usually leads to the conventional $J=J_c \sin{\phi}$
relation, except for a special magnetic configuration with mutually
perpendicular directions where a
transition between "0" and "$\pi$" states occurs \cite{B-N}.
Then the first harmonic vanishes and the current is given by the
second harmonic $\sim \sin 2\phi$; however, its amplitude is relatively small.
In general, Josephson junctions exhibiting double-periodic behavior,
besides being interesting objects in proximity-effect studies,
may be useful in flux-qubit design schemes \cite{Ioffe}.
Josephson frequency  doubling was predicted  in other types of
junctions involving unconventional superconductors, such as s-p
\cite{Asano}, s-d-s \cite{Ioffe}, p-p and d-d junctions
with specific misorientation angles of the order parameter \cite{Yokoyama}.
It was observed in experiments involving
d-wave grain-boundary junctions \cite{Il'ichev}.
It should be stressed, however, that in all these cases the frequency doubling
occurs at isolated points in the parameter space where the first
harmonic vanishes. Moreover, the magnitude of the current is
suppressed in comparison with the usual value
$\sim \Delta /e R_n$, where $R_n$ is the normal-state resistance.

In this work, we consider the magnetic exchange interaction induced by
Josephson currents in a dirty S/F heterostructure exhibiting TPE. We
show that this interaction may prefer non-collinear
magnetic configurations and the preferred direction depends
continuously on the superconducting phase
difference. Thus,  the static magnetic configuration can be controlled by
the applied phase difference. We then consider the influence
of feedback from the magnetic moments on the AC Josephson
effect. The magnetic system exhibits a range of different
behaviors, from  simple harmonic oscillations to fractional-frequency
periodic behavior and chaotic motion.  A finite
zero-frequency deviation from the equilibrium configuration may appear,
allowing  control of the direction of the average magnetization
also by an applied voltage. The magnetic feedback complicates the
behavior of the current in the time domain, making it generally
impossible to express it in terms of a current-phase relation.
On the other hand, we find that both in the low- and high frequency
limit such a relation becomes meaningful, with the current exhibiting a
double-phase  dependence, $J\sim \sin 2\phi(t)$ or
$J\sim \cos 2\phi(t)$. The critical current in the
low-frequency regime is of the order of the
value $ E_{Th}/e R_n$, characteristic for diffusive systems.
The unusual cosine dependence of the Josephson
current appears when Gilbert damping is important in the magnetic
dynamics, breaking the time-reversal symmetry. At high frequencies,
the magnetization cannot effectively follow the phase variation,
leading to a $\sim 1/\omega^2$ suppression of the effective Josephson
coupling. At even higher frequencies, the damping is
dominant, and the frequency dependence becomes $\sim 1/\omega$. The
presence of damping is expressed in the appearance of a dc component
of the current leading to a finite resistance.

\begin{figure}[b]
\begin{center}
\includegraphics[width=0.4\textwidth]{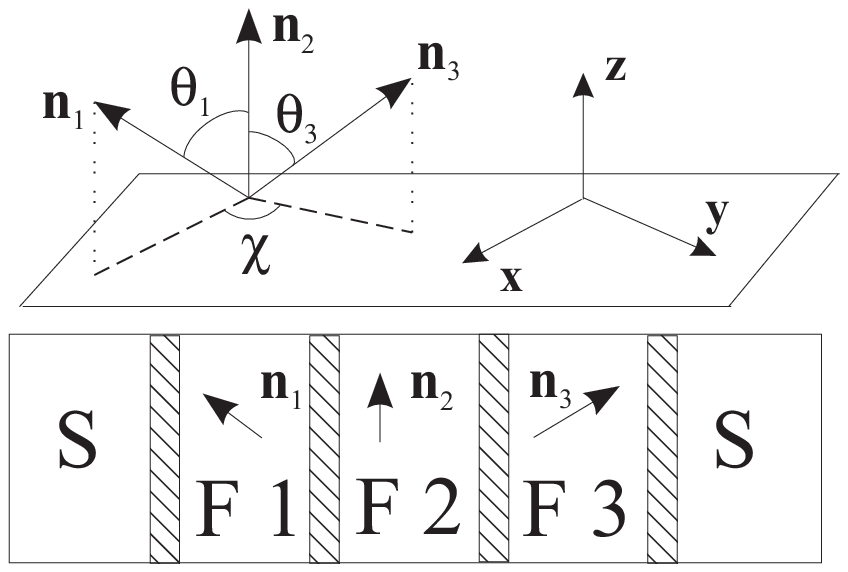}
\end{center}
\caption{The experimental setup}
\label{fig:setup}
\end{figure}

{\bf The system}. We consider an S/F heterostructure described in
Fig.~\ref{fig:setup} which is a minimal discrete setup
exhibiting the triplet proximity effect. Two magnetic regions 1 and 3 are
adjacent to the superconducting reservoirs that induce proximity
mini-gaps $\Delta_{1,3}$ in them.
Between these regions there is an additional  magnetic region 2
whose length is much larger than $\xi_h$ and where triplet
superconducting correlations are induced.
This region is assumed to be weakly polarized (metallic), so that both spin directions are
present at the Fermi surface.
The magnetic regions are characterized by the exchange
energies $h_i$, while the magnetization directions $\bm{n}_{i}$ are
specified by the angles $\theta_1$,
$\theta_3$ and $\chi$ as shown in Fig.~\ref{fig:setup}.
Assuming that the conductances of these regions are much higher than
the conductances $g^n_{1,3}$ of the connectors between them, our
system can be described by  a circuit-theory  model for the triplet
proximity effect used in Ref. \onlinecite{B-N}.
Magnetization directions of regions 1 and 2 are assumed fixed, {\em e.g.} by
pinning to an antiferromagnetic substrate, or by geometrical shaping,
with the angle between them being $\theta_1$. On the other hand,
magnetization $\bm{n}_3$ is free to rotate, with region 3  separated
by a normal spacer from region 2 in order to avoid exchange coupling
between them.

In accordance with the model assumptions, regions 1 and 3 act as
effective S-F reservoirs, hence their energies are independent of the
magnetic configuration. On the other hand, triplet superconducting
correlations extending through region 2, are very sensitive to the
magnetization directions. Hence the configuration-dependent part of
the energy can be found by integrating over the density
of states (DOS) in region 2. The DOS for each spin direction  is given
by \cite{B-N}
\beq
  \nu^{\uparrow,\downarrow}(\varepsilon) = \frac{\nu_0}{2}
{\rm Re}
\left( 1- \frac {a_1^2+a_3^2 + 2a_1 a_3 \cos(\phi \pm \chi)}
{ (b_1+b_3-i\e/E_{Th})^2}
\right)^{-\frac{1}{2}},
\label{eq:nuupdown}
\eeq
where $ \nu_0$ is the normal-state DOS,
$a_k= g^n_k|\Delta_k|\sin\theta_k/(g^n_1+g^n_3)\sqrt{h^2_k -|\Delta_k|^2}$,
$b_k=g^n_k h_k/(g^n_1+g^n_3)\sqrt{h^2_k -|\Delta_k|^2}$, $\phi$ is the
superconducting phase difference, and $E_{Th}$ is the Thouless energy of the
structure. Using this expression, one can  see that the energy is
given by a logarithmic integral and the main contribution comes from
$\epsilon \gg E_{Th}$. In the leading order one obtains
\beq \label{eq:energy}
  E=\frac{\nu_0 v_2}{2} \log \frac{\Delta_{cut}}{E_{Th}}
  E_{Th}^2\left(a_1^2+a_3^2+2 a_1 a_3 \cos \phi
  \cos \chi \right ) \ ,
\eeq
where $v_2$ is the volume of the magnetic region 2 and $\Delta_{cut} \simeq
\textrm{min}(\Delta_i, h_i-\Delta_i)$
is a cutoff energy.  This
expression can be written in a form presenting explicitly the
dependence on the orientation angles $\theta_3$ and $\chi$,
\beq
  E=p_3^2 \sin^2 \theta_3+2 p_1 p_3 \sin \theta_3 \cos \phi \,\cos \chi,
\eeq
with $p_{1,3}$ being effective exchange couplings for the magnetic vector $\bm{n}_3$.
The stable configuration is achieved when all magnetization directions
are in the same
plane, denoted in the following as the $\bm{x}-\bm{z}$ plane, and
$\bm{n}_3$ is tilted with respect to $\bm{n}_2$ by a
finite angle satisfying
\beq \label{eq:stable}
   \sin \theta_3 = \frac{p_1}{p_3} |\cos \phi| \ .
\eeq
This angle depends continuously on the applied superconducting phase
difference $\phi$, while the angle $\chi$ assumes the values $0$ or
$\pi$ so that the product $ \cos \phi \cos \chi$ is negative.
In fact, there are two stable directions, given by the angles
$\theta_3$ and $\pi-\theta_3$.
In what follows we will treat them as equivalent, since they
correspond to the same current. The energy of the stable configuration
is given by
\beq
 E_{min}=-p_1^2    \cos^2 \phi \ .
\eeq
Hence allowing the magnetization direction $\bm{n}_3$ to orient itself
along the stable direction leads to the current-phase relation $J =
J_c \sin 2\phi$.

{\bf Low frequencies}. When a small voltage $V$ is applied
to the structure, such that the corresponding frequency $\omega_J=2eV/\hbar$
is much smaller than the characteristic frequency of the magnetic
system $\omega_m$ (see below),  the vector $\bm{n}_3$ follows the stable direction given by
Eq. (\ref{eq:stable}), performing slow oscillations in the
$\bm{x}-\bm{z}$ plane. The alternating  Josephson current oscillates
with the double frequency
\beq
 J=\frac{2 e}{\hbar} p_1^2 \sin \frac{4 e V}{\hbar} t \ ,
\eeq
while the critical current remains of the same order of magnitude as
in the case with a fixed magnetic configuration.

\begin{figure}[t]
\begin{center}
\makebox[ \textwidth][l]{
\hspace{-0.0\textwidth}
\includegraphics[width=0.25\textwidth]{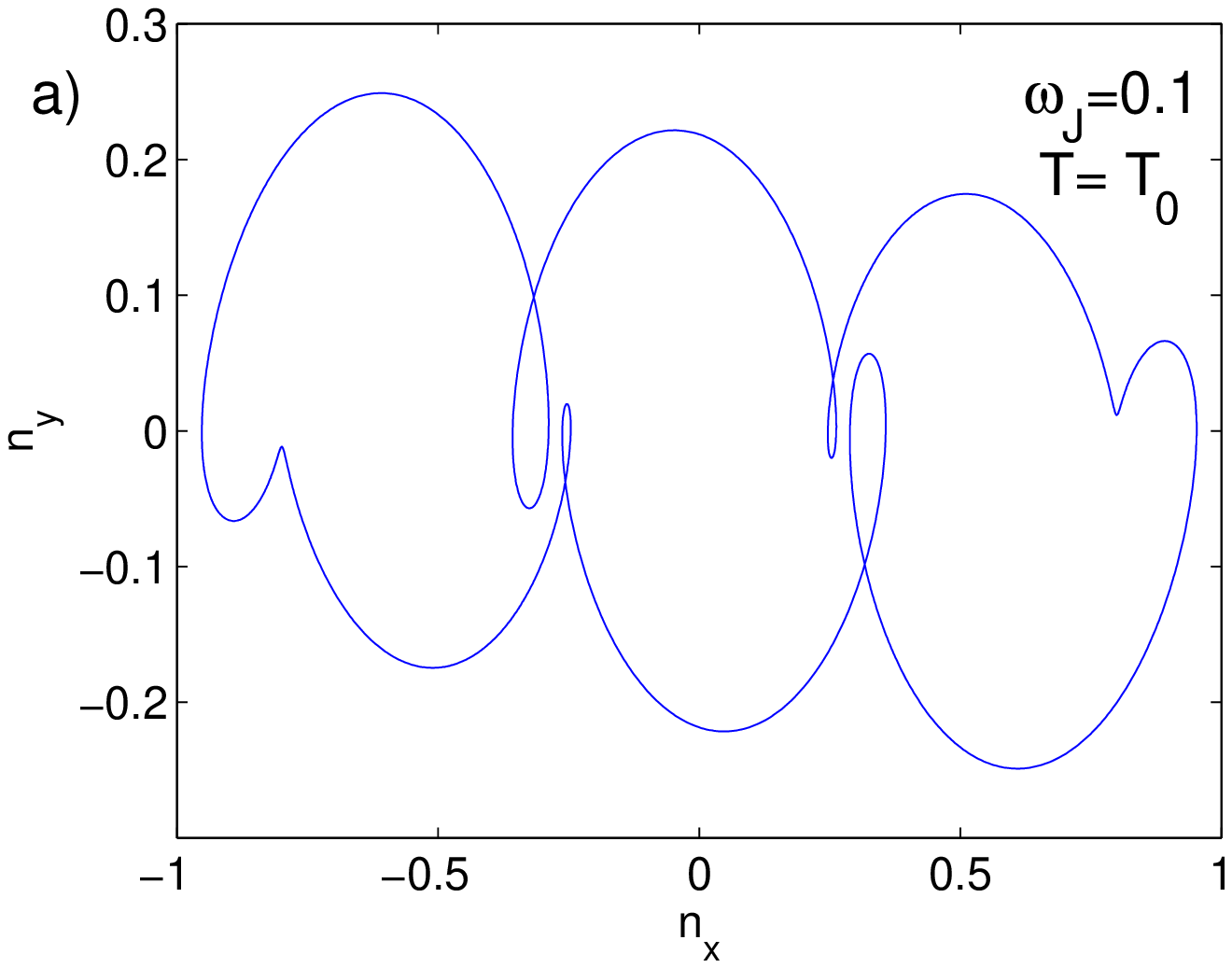}
\hspace{-0.03\textwidth}
\includegraphics[width=0.25\textwidth]{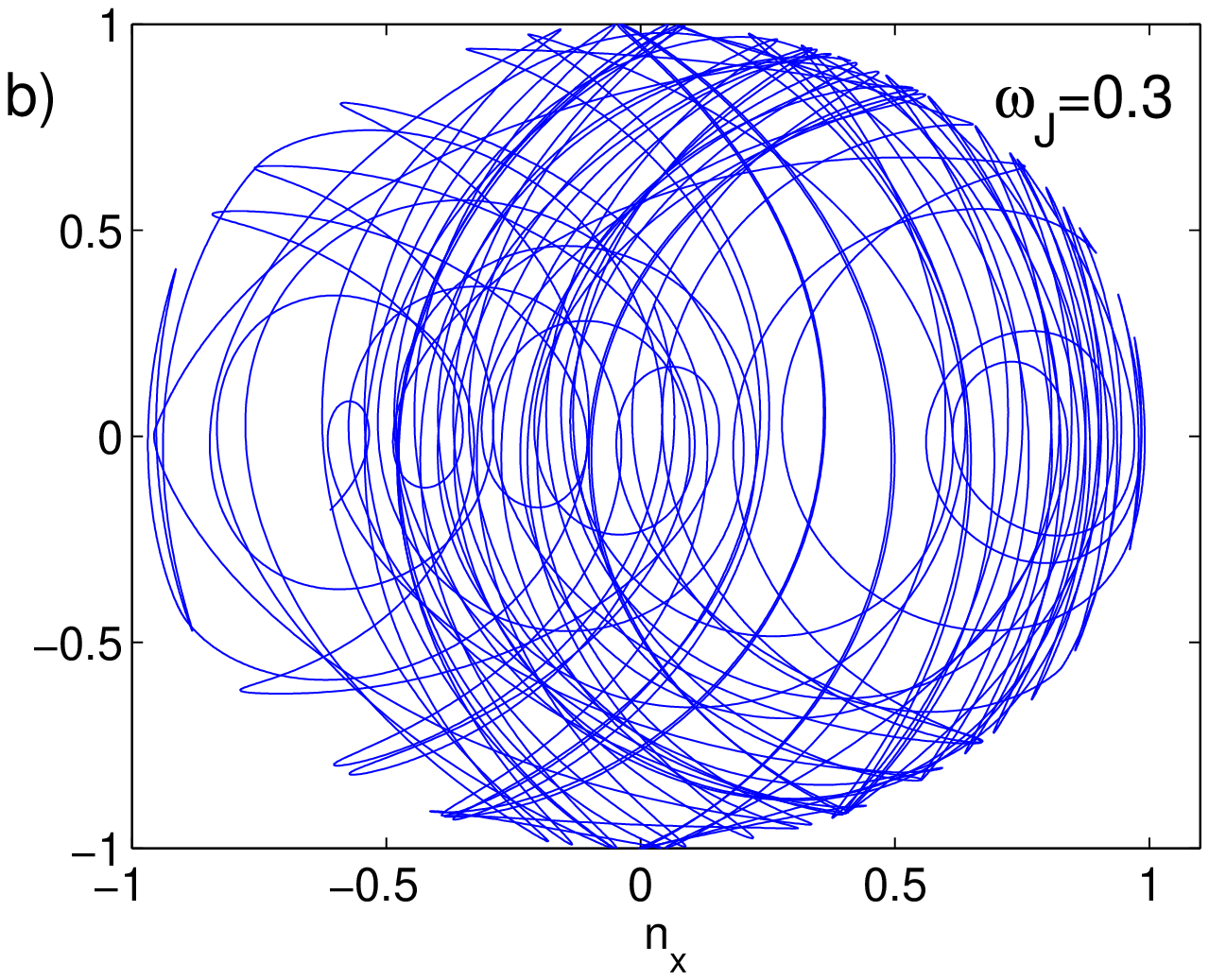}
}
\makebox[ \textwidth][l]{
\hspace{-0.01\textwidth}
\includegraphics[width=0.25\textwidth]{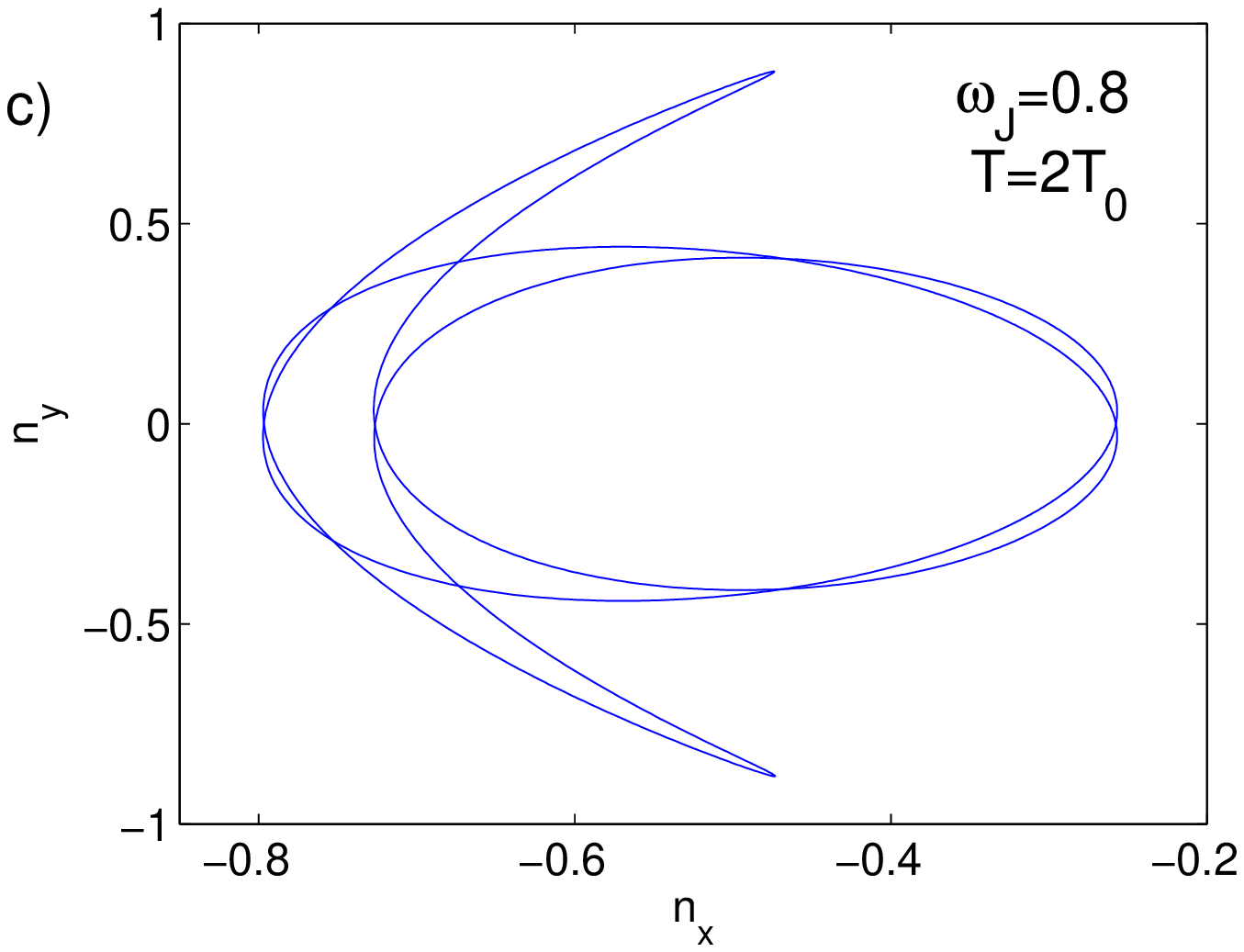}
\hspace{-0.03\textwidth}
\includegraphics[width=0.25\textwidth]{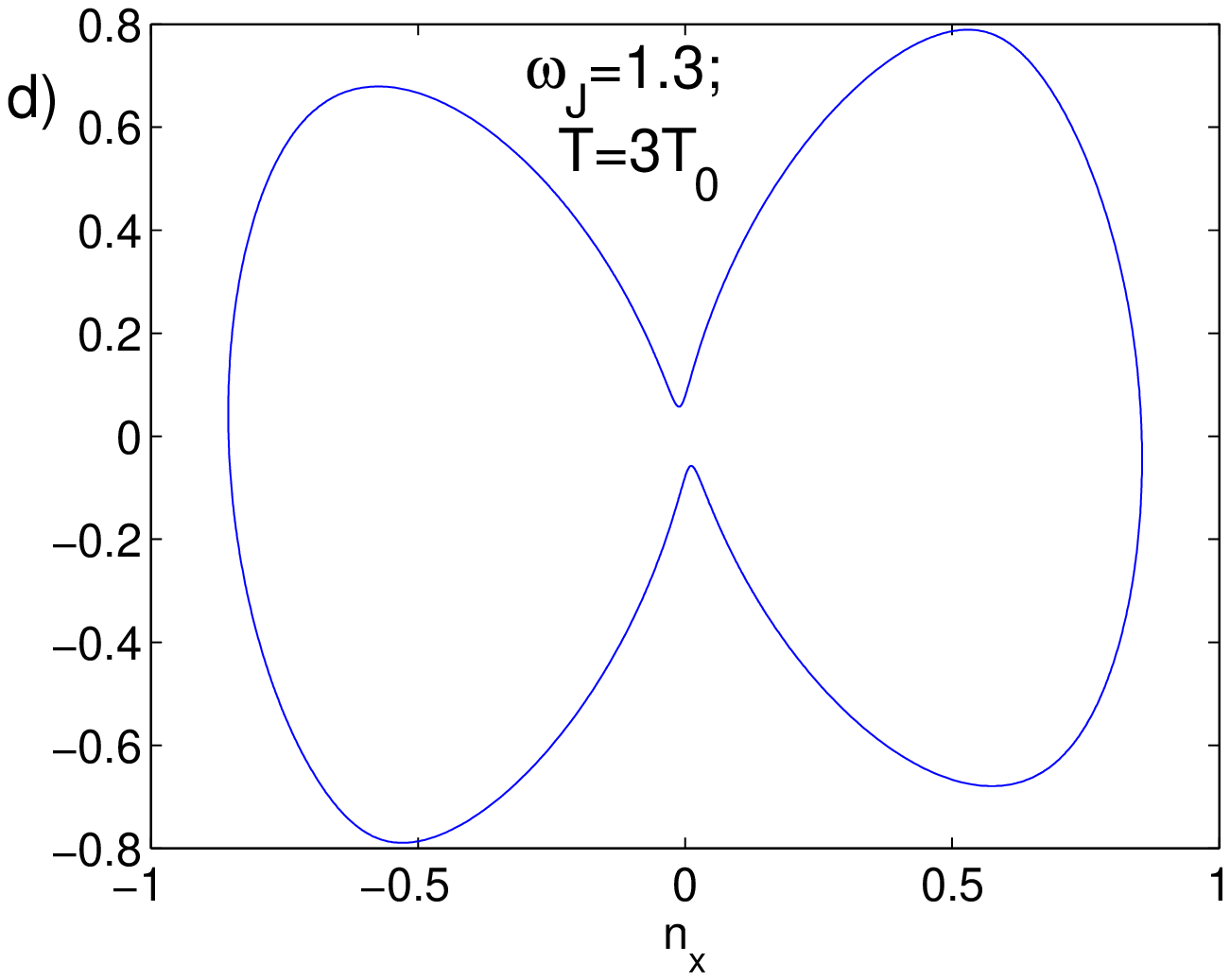}
}
\end{center}
\caption{Trajectories of the magnetization vector in the $x-y$ plane
for different Josephson
frequencies (given in units of $\omega_m$). Trajectory (b) is chaotic, while trajectory (c) has a
finite zero-frequency component
for $n_x$. Here $T$ is the period of the trajectory, and
$T_0=2\pi/\omega_J$. For comparison, the
low-frequency trajectory lies entirely on the $\bm{x}$ axis. }
\label{fig:traj}
\end{figure}

For higher Josephson frequencies, the variation of $\bm{n}_3$ is no
more limited to the $\bm{x}-\bm{z}$ plane. Instead, the magnetization
performs a variety of non-harmonic motions whose frequency may be a
multiple or a fraction of the driving frequency $\omega_J$
[Fig.~\ref{fig:traj} (a), (c), (d)]. For certain trajectories
the time average of $\theta_3$ is finite [Fig.~\ref{fig:traj} (c)],
corresponding to a tilt of $\bm{n}_3$ away from the equilibrium in
response to an applied voltage. Within some frequency intervals, the
motion is chaotic, as shown in Fig.~\ref{fig:traj} (b). In these
intermediate regimes, the Josephson current shows a complicated time
dependence which is generally not periodic in $2\pi/\omega_J$. Hence
this dependence cannot be parameterized in terms of the phase. Instead,
one can speak of a Josephson current with a time-dependent coupling.

{\bf High frequencies}.
When applied voltage is high, the Josephson frequency becomes
much higher than the magnetic frequencies.
In this case the magnetic vector $\bm{n}_3$ cannot effectively follow
the fast oscillations of
the potential, and the time-averaged potential seen by  $\bm{n}_3$ has a minimum
for $\bm{n}_3 \parallel \bm{z}$. The motion of $\bm{n}_3$ can be
determined by  expanding $\bm{n}_3=\bm{z}+\delta \bm{n}$ and using
a linearized Landau-Lifshits-Gilbert (LLG) equation,
\beq  \label{eq:LLG}
  \delta \dot{\bm{n}}= \bm{z} (-\gamma \times \bm{H}_{eff}+\alpha
  \delta \dot{\bm{n}}) \ ,
\eeq
where $\gamma$ is the gyromagnetic ratio, $\alpha$ is the effective damping
coefficient, $\bm{H}_{eff}=-\partial E/\partial \bm{m}_3$ is the
effective field, and $\bm{m}_3$ is the magnetization density of
magnetic region 3.

\begin{figure}[b t]
\begin{center}
\includegraphics[width=0.45\textwidth]{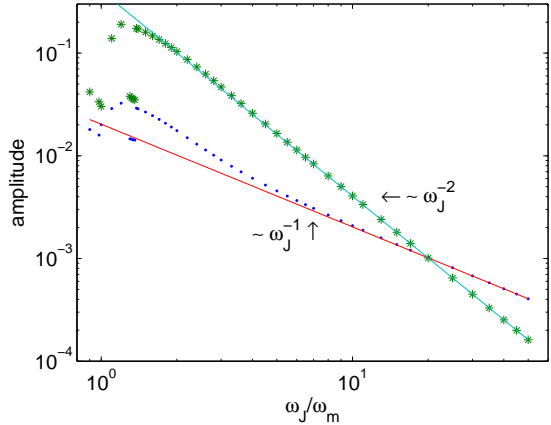}
\end{center}
\caption{The absolute value of the Josephson current harmonics
proportional to $\sin 2 \omega_J t$ (asterisks) and to
$1-\cos 2 \omega_J t$ (dots). Solid lines are fits $\sim 1/\omega$ and
$1/\omega^2$. The data points are obtained from numerical integration of the
full (non-linear) LLG equation.}
\label{fig:harmonics}
\end{figure}

When Gilbert damping is negligible, the trajectory of $\bm{n}_3$ has a
very low aspect ratio, so that the motion is almost completely
confined to the $\bm{y}$ axis. It is given by
\begin{eqnarray}
  \delta n_x&=& \frac{\gamma^2 p_1 p_3^3 \hbar^2}{ e^2 V^2 m_3^2 }\cos
   \frac{2 e V t}{\hbar} \ ;
   \nonumber \\
  \delta n_y&=&\frac{\gamma p_1 p_3 \hbar}{ e V m_3 } \sin \frac{2 e V
   t}{\hbar} \ .
\end{eqnarray}
Thus at high frequencies, $\bm{n}_3$ precesses in phase with the
voltage pumping. This leads to an increase in the Josephson energy,
and, correspondingly, a negative Josephson current,
\beq
J=-\frac{2 \hbar}{e} \left( \frac{ \gamma p_1 p_3^2  }{ V m_3 } \right )^2 \sin
  \frac{4 e V t}{\hbar} \ .
\eeq
Hence in the high-frequency regime the system shows not only frequency
doubling, but also an effective $\pi$-junction behavior. The magnitude
of the current is suppressed as $\sim V^{-2}$ as shown in
Fig.~\ref{fig:harmonics}.

The neglect of damping is justified as long as $\alpha \omega_J \ll
\omega_m=\gamma p_3^2/m_3$. When the voltage is high enough, this
condition is not satisfied anymore, and the dissipation starts to be
important. As the Josephson frequency becomes so large that the
opposite inequality holds, the motion of $\bm{n}_3$ is determined by
the driving against the damping force,
\beq
  \delta \bm{n} = \frac{ \gamma p_1 p_3 \hbar }{e V m_3 } (-\alpha
  \hat{\bm{x}}+\hat{\bm{y}} ) \sin \frac{2 e V t}{\hbar} \ .
\eeq
Then the Josephson current is given by
\beq
  J= \frac{2 \alpha \gamma p_1^2 p_3^2 }{m_3  V} \left(1-\cos \frac{4
    e V t}{\hbar}\right) \ .
\eeq
Note the unusual cosine dependence on the phase. It occurs since the
time-reversal symmetry is broken by the dissipation in this
regime. Due to the same reason, a zero-frequency component
of the current appears, signifying the onset of a finite nonlinear
resistance across the structure.
Since this regime is governed by the damping,
the current amplitude is proportional to $\alpha$, while the
suppression
$\sim 1/V$ is weaker in this regime (Fig.~\ref{fig:harmonics}).

To estimate the magnetic dynamics frequency $\omega_m$, we use typical
values $E_{Th}\sim 1$ meV, $\nu_0\sim 1/( \mathrm{eV/atom})$, and
$m_3\sim 1 \mu_B/\mathrm{atom}$, where $\mu_B$ is the Bohr
magneton. Then $\omega_m\sim v_2/v_3$ GHz, where $v_{2,3}$ are the
volumes of the corresponding magnetic regions. As this frequency is
quite low, observation of the high-frequency regimes should present no
difficulty. On the other hand, the low-frequency AC regime would
require extremely low voltages, below 1 $\mu$V. A reasonable
alternative would be incorporating the structure in a superconducting
loop and measuring the  Josephson current as a function of the applied
flux.

Applicability of our model requires that any magnetic anisotropy
of part 3 should be smaller than the proximity-induced energy,
Eq. (\ref{eq:energy}). With the above values of the parameters
it is of the order of $10^4 \times v_2$ J/m$^3$, so one should
choose materials with low value of the cristalline anisotropy,
such as permalloy.
 Finally, we emphasize that
the properties discussed above
are specific for metallic systems. In half-metals, the behavior will
be very different. Thus, in the low-frequency regime $\bm{n}_3$
precesses around $\bm{n}_2$ at a constant angle $\theta_3$, while the
Josephson current vanishes.

{\bf Conclusions}. We have considered the AC Josephson effect in a
S/F/S structure with magnetic dynamics coupled to the dynamics of
superconducting correlations. The magnetic configuration in the
structure was assumed to be non-uniform so that the structure exhibits a
triplet proximity effect. Variation of the magnetic configuration is
shown to essentially modify the current behavior that can be observed
in the appearance of fractional Shapiro steps. Thus measurement of the
Josephson current would provide information about the coupling and
self-consistent feedback dynamics between the superconducting and
magnetic degrees of freedom. The coupling also allows to control the
magnetization direction by means of applied voltage or superconducting
phase. In the low-frequency limit, the magnetization follows the
immediate potential minimum, leading to a $\sim \sin 2\phi$
current-phase relation. The critical current has the same order of
magnitude  $E_{Th}/e R_n$
as that due to the usual
singlet proximity effect in dirty structures. In the high-frequency
regime, as long as the damping is not important, the Josephson current
is negative, corresponding to a $\pi$-junction behavior. It is
suppressed by a factor $\sim (\omega_m/\omega_J)^2$ relative to the
low-frequency regime. At even higher frequencies, Gilbert damping starts
playing the major role in the dynamics. Then the time-reversal
symmetry is broken and the current-phase relation takes an unusual
cosine form. In addition, a DC component of the current appears,
manifesting itself in a finite resistance. The current suppression
becomes weaker in this regime.

The authors are grateful to G.~E.~W.~Bauer and Yu.~V.~Nazarov for
useful discussions.
This work was supported by EC Grant No. NMP2-CT2003-505587 (SFINX).

\end{document}